\documentclass[%
  a4paper, twocolumn, pra, superscriptaddress, amsmath, amssymb, floatfix
]{revtex4-1}
\usepackage{graphicx}
\usepackage{epstopdf}
\usepackage[table]{xcolor}
\usepackage{braket}
\usepackage{algorithm}
\usepackage{algpseudocode}
\usepackage[colorlinks,bookmarksopen=false]{hyperref}
\usepackage[caption=false]{subfig}

\hypersetup{%
  pdfstartview={FitH},
  linkcolor=blue,
  citecolor=blue,
  filecolor=black,
  menucolor=black,
  urlcolor =black
}

\newcommand{\RED}[1]{\textcolor{black}{#1}}

\newcommand{\im}{\ensuremath{\textup{i}}}

\newcommand{\rp}{\ensuremath{\mathfrak{Re}}}
\newcommand{\ip}{\ensuremath{\mathfrak{Im}}}

\newcommand{\op}[1]{\ensuremath{\boldsymbol{\mathsf{\hat{#1}}}}}

\newcommand{\lop}[1]{\ensuremath{\boldsymbol{\mathsf{\mathcal{#1}}}}}

\newcommand{\uop}{\op{1}}

\newcommand{\tr}[3]{\ensuremath{\textup{Tr}_{#1}^{#2}\left[#3\right]}}

\newcommand{\pulse}{\varepsilon}
\newcommand{\error}{\epsilon}

\begin{document}

\title{%
  Beating the limits with initial correlations
}

\author{Daniel Basilewitsch}
\affiliation{Theoretische Physik, Universit\"{a}t Kassel, D-34132 Kassel,
Germany}

\author{Rebecca Schmidt}
\affiliation{Turku Centre for Quantum Physics, Department of Physics and
Astronomy, University of Turku, FIN-20014 Turku, Finland}
\affiliation{Center for Quantum Engineering, Department of Applied Physics,
Aalto University School of Science, P.O. Box 11000, FIN-00076 Aalto, Finland}

\author{Dominique Sugny}
\affiliation{Laboratoire Interdisciplinaire Carnot de Bourgogne,
(ICB), UMR 5209 CNRS-Universit\'e de Bourgogne Franche Comt\'e, 9
Av. A. Savary, BP 47 870, F-21078 DIJON Cedex, France}
\affiliation{Institute for Advanced Study, Technische Universit\"at
M\"unchen, Lichtenbergstrasse 2 a, D-85748 Garching, Germany}

\author{Sabrina Maniscalco}
\affiliation{Turku Centre for Quantum Physics, Department of Physics and
Astronomy, University of Turku, FIN-20014 Turku, Finland}
\affiliation{Center for Quantum Engineering, Department of Applied Physics,
Aalto University School of Science, P.O. Box 11000, FIN-00076 Aalto, Finland}

\author{Christiane P. Koch}
\affiliation{Theoretische Physik, Universit\"{a}t Kassel, D-34132 Kassel,
Germany}
\email{christiane.koch@uni-kassel.de}

\date{\today}

\begin{abstract}
  Fast and reliable reset of a qubit is a key prerequisite for any quantum
  technology. For real world open quantum systems undergoing non-Markovian
  dynamics, reset implies not only purification, but in particular erasure of
  initial correlations between qubit and environment. Here, we derive optimal
  reset protocols using a combination of geometric and numerical control theory.
  For factorizing initial states, we find a lower limit for the entropy
  reduction of the qubit as well as a speed limit. The time-optimal solution is
  determined by the maximum coupling strength. Initial correlations, remarkably,
  allow for faster reset and smaller errors. Entanglement is not necessary.
\end{abstract}

\maketitle

\section{Introduction} \label{sec:intro}
Quantum technology requires re-usable qubits~\cite{DiVincenzo2000}.  A reliable
reset to a well-defined state is \RED{therefore} vital.  \RED{This is true} no
matter \RED{whether} the quantum system in question is to be used repeatedly,
\RED{as} in the case of quantum computing~\cite{Fernandez2004, Ladd2010,
Reed2010, Ristle2012, Johnson2012, Govia2015}, or \RED{whether} a cycle is to be
performed, as required for quantum thermodynamical machines~\cite{Kosloff2014,
Gelbwaser2015, Rossnagel2016, Karimi2016, Watanabe2017}. \RED{Reset} implies
purification or cooling~\cite{Gerling2013, Horowitz2014, Pietro2016,
Tuorila2016}, since quantum systems are inevitably in contact with their
environment. The corresponding entropy reduction can be achieved in two
ways---by employing an auxiliary degree of freedom with lower entropy than the
system for an entropy swap~\cite{Horowitz2014, Pietro2016} or by coupling the
system to a reservoir where the steady state coincides with the desired reset.
The relaxation in the latter case is typically sped up by extra
means~\cite{Gerling2013, Tuorila2016}, \RED{which is important since} fast
protocols are desirable for error prevention. In both settings for cooling, the
\RED{coupling to the }entropy sink, i.e., the environment, can be switched on
and off at will.

Cooling \RED{alone} is not enough for a complete reset \RED{which} also requires
the erasure of any correlations between system and environment. This aspect is
typically not taken into account, due to the assumption of weak coupling
\RED{between system and environment in standard models}. However,
\RED{persistent correlations may affect the functioning of a quantum device. For
example,} different cycles of a quantum heat engine do not show the same
performance in the presence of intercycle coherence~\cite{Watanabe2017}. In
general, the assumption of negligible correlations is hardly justified in
mesoscopic devices such as superconducting qubits~\cite{weiss2012}. \RED{These
systems are also known for their} non-Markovian dynamics, \RED{displaying memory
effects due to the coupling to the environment}.

Here, we focus on the role of initial correlations between system and
environment for qubit reset. Using quantum optimal control, we show that initial
correlations can not only be erased, but turn out to be an asset for
purification. With initial correlations, we are able to outperform the best
possible uncorrelated reset protocol both in fidelity and minimal time. Our
results suggest to actively exploit initial correlations between system and
environment in quantum technology.

\RED{In more detail}, we consider a qubit in contact with an environment which
gives rise to non-factorizing dynamics. Assuming the qubit was used in a quantum
computation or \RED{in} a thermodynamic cycle, the task is to erase the
correlations with the environment and transfer the qubit into a well-defined
pure state. \RED{In other words}, we aim at cooling the qubit \emph{below} the
steady state of the open system and, at the same time, erase all correlations.
To this end, we employ quantum optimal control theory~\cite{Glaser.EPJD.69.279}.
By definition, only the system, i.e., the qubit, is controllable\RED{;} the
environment and the system-environment coupling are not. We also investigate
whether entanglement and memory effects facilitate qubit reset. \RED{This is
motivated by recent} evidence that non-Markovian dynamics might be a resource
for control tasks such as cooling~\cite{RS2011} or gate
implementation~\cite{Rebentrost.PRL.102.090401, Reich2015}.

The paper is organized as follows. Section~\ref{sec:model} introduces the model
we study. The numerical results for optimal qubit reset are presented in
section~\ref{sec:num}. The control problem can be solved analytically in certain
limits, as shown in section~\ref{sec:ana}. The analytical results provide an
intuitive interpretation of the reset protocols obtained numerically.
Section~\ref{sec:conclusions} concludes.

\section{Model} \label{sec:model}
\begin{figure}[tb]
  \centering
  \includegraphics{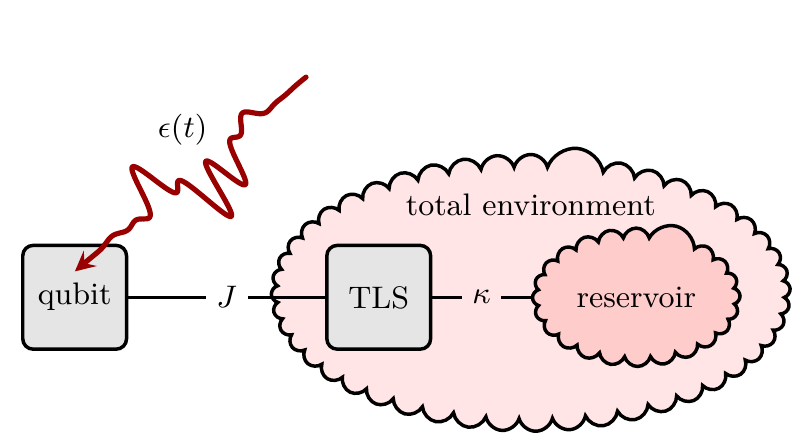}
    \caption{%
      We consider a qubit strongly coupled to a two-level system (TLS) that is
      weakly coupled to a reservoir. Together, TLS and reservoir define the
      total environment for the qubit. Due to the strong coupling $J$ between
      qubit and TLS, the qubit dynamics may become non-Markovian. The
      TLS-reservoir couples with strength $\kappa$. The coupling between qubit
      and the reservoir is only indirect.
    }
    \label{fig:model:model}
\end{figure}
Our system consists of a qubit in interaction with an external field. The
Hamiltonian reads
\begin{equation} \label{eq:model:hamQ}
  \op{H}_{\text{Q}}(t)
  =
  -\frac{\hbar \omega_{\text{Q}}}{2}\op{\sigma}^{z}_{\text{Q}}
  -\frac{\hbar \pulse(t)}{2}\op{\sigma}^{z}_{\text{Q}}.
\end{equation}
Here, $\omega_{\text{Q}}$ is the qubit's level splitting and $\pulse(t)$
a control field, to be determined by optimal control theory. $\op{\sigma}^{i}$,
$i=\{x,y,z\}$, are the usual Pauli matrices.

This single qubit is coupled to an environment that may, in general, give rise
to non-Markovian dynamics. Such an environment can be mapped onto a pseudo-mode
weakly coupled to a large bath of harmonic modes~\cite{Garraway1996}, as
depicted in Fig.~\ref{fig:model:model}. The pseudo-mode, which acts as a memory,
is taken to also be a two-level system (TLS), with Hamiltonian
$\op{H}_{\text{TLS}} = - \frac{\hbar \omega_{\text{TLS}}}{2}
\op{\sigma}^{z}_{\text{TLS}}$ and level splitting $\omega_{\text{TLS}}$. The
pseudo-mode is not necessarily weakly coupled to the system qubit. We therefore
treat the interaction between the system qubit and the memory TLS exactly. This
allows to fully capture the correlations we are interested in. For the rest of
the environment, we employ the usual approximations,  leading to the standard
Markovian master equation for the joint state $\op{\rho}(t)$ of qubit and
TLS~\cite{Garraway1996, Rebentrost.PRL.102.090401,Addis2014,Reich2015},
\begin{equation} \label{eq:model:LvN}
  \begin{aligned}
    \im \hbar \frac{\text{d}}{\text{d}t} \op{\rho}(t)
    &=
    \left[\op{H}(t), \op{\rho}(t)\right]
    +
    \lop{L}_{D}\left[\op{\rho}(t)\right],
    \\
    \lop{L}_{D}\left[\op{\rho}(t)\right]
    &=
    \im \hbar \sum_{k=1,2} \kappa \left(%
        \op{L}_{k} \op{\rho}(t) \op{L}_{k}^{\dagger}
        - \frac{1}{2} \left\{%
          \op{L}_{k}^{\dagger} \op{L}_{k}, \op{\rho}(t)
        \right\}
      \right),
  \end{aligned}
\end{equation}
with Hamiltonian
$\op{H}(t) = \op{H}_{\text{Q}}(t) \otimes \uop_{\text{TLS}} + \uop_{\text{Q}}
\otimes \op{H}_{\text{TLS}} + \op{H}_{\text{int}}$. The interaction between
qubit and TLS given by
\begin{equation} \label{eq:model:hamQT}
  \begin{aligned}
    \op{H}_{\text{int}}
    &=
    J \left(%
          \op{\sigma}^{x}_{\text{Q}} \otimes \op{\sigma}^{x}_{\text{TLS}}
      \right).
  \end{aligned}
\end{equation}
The Lindblad operators $\op{L}_{k}$ model the thermal equilibration between the
TLS and the remaining reservoir and correspond to those of the optical master
equation~\cite{Breuer.book}, $\op{L}_{1} = \sqrt{N + 1} (\uop_{\text{Q}} \otimes
\op{\sigma}_{\text{TLS}}^{-})$, $\op{L}_{2} = \sqrt{N} (\uop_{\text{Q}} \otimes
\op{\sigma}_{\text{TLS}}^{+})$, where $N = 1/\left(e^{\beta \hbar
\omega_{\text{TLS}}} - 1\right)$ and $\beta$ is the inverse thermal energy of
the reservoir. The state of the qubit is obtained by tracing out the degrees of
freedom of the memory TLS at each instant in time,
$\op{\rho}_Q(t)=\tr{\text{TLS}}{}{\op{\rho}(t)}$.

\RED{Cooling requires population relaxation. This motivates our choice of
exchange interaction between qubit and memory TLS in
Eq.~\eqref{eq:model:hamQT}.}
We take the coupling $J$ between qubit and TLS to be larger than the coupling
$\kappa$ (otherwise the dynamics of the qubit would be Markovian). On the other
hand, $J$ is still small with respect to the level splitting
$\omega_{\text{TLS}}$ of the TLS\@. The corresponding timescale separation
ensures detailed balance and accord with the second law of
thermodynamics~\cite{Levy.EPL.107.20004}. Note that $\kappa$ refers to a rate,
in a physical sense, rather than a coupling strength but, since both can't be
distinguished mathematically, we refer to it as a coupling.

We will analyze several initial states for the qubit and memory TLS\@. Since the
TLS is part of the environment, we always assume it to be initially in thermal
equilibrium with the reservoir. To fully understand the role of initial
correlations, we start from the factorized case and then generalize it. For the
sake of comparability, we assume that the initial state of the qubit is
quasi-thermalized with the reservoir as well. Their respective initial states
read
\begin{align} \label{eq:model:state_two_level_therm_eq}
  \op{\rho}_{\alpha}^{\text{th}}
  =
  \frac{e^{x_{\alpha}}\Ket{0}\Bra{0} + e^{-x_{\alpha}}\Ket{1}\Bra{1}}
  {2 \cosh\left(x_{\alpha}\right)},
  \qquad
  x_{\alpha}
  =
  \frac{\hbar\omega_{\alpha} \beta}{2},
\end{align}
with $\alpha \in \left\{\text{Q}, \text{TLS}\right\}$. In the factorized case,
the joint state of qubit and TLS at $t=0$ reads
\begin{align} \label{eq:model:state_fact_init}
  \op{\rho}^{\text{init}}_1
  =
  \op{\rho}_{\text{Q}}^{\text{th}} \otimes \op{\rho}_{\text{TLS}}^{\text{th}}.
\end{align}
For non-factorizing initial conditions, we first investigate the fully
thermalized state of qubit and TLS, which is the steady state and therefore
a natural choice. It reads
\begin{align} \label{eq:model:th_bipartite}
  \op{\rho}^{\text{init}}_2
  =
  \op{\rho}^{\text{th}}
  =
  \frac{1}{Z}
  \begin{pmatrix}
    \lambda_{+} & 0 & 0 & \zeta_{+}
    \\
    0 & \mu_{+} & \zeta_{-} & 0
    \\
    0 & \zeta_{-} & \mu_{-} & 0
    \\
    \zeta_{+} & 0 & 0 & \lambda_{-}
  \end{pmatrix},
\end{align}
where $Z = 2 \left[\cosh\left(x_{+}\right) + \cosh\left(x_{-}\right)\right]$ is
the partition function and
\begin{equation}
  \begin{alignedat}{3}
    \delta_{\pm}
    &=
    \omega_{\text{Q}} \pm \omega_{\text{TLS}},
    \qquad
    &&\lambda_{\pm}
    &&=
    \cosh\left(x_{+}\right) \pm \frac{\delta_{+}}{\Omega_{+}}
    \sinh\left(x_{+}\right),
    \\
    x_{\pm}
    &=
    \frac{\Omega_{\pm} \beta}{2},
    \qquad
    &&\mu_{\pm}
    &&=
    \cosh\left(x_{-}\right) \pm \frac{\delta_{-}}{\Omega_{-}}
    \sinh\left(x_{-}\right),
    \\
    \Omega_{\pm}
    &=
    \sqrt{\delta_{\pm}^{2} + 4 J^{2}},
    \qquad
    &&\zeta_{\pm}
    &&=
    - \frac{2 J}{\Omega_{\pm}} \sinh\left(x_{\pm}\right).
  \end{alignedat}
\end{equation}
Since this state can always be obtained by waiting (or speeding up of
thermalisation), the control problem in general is solved, if we can solve it
for the steady state. For the chosen parameters~\cite{Karimi2016}, the initial
correlations of the thermalized state are rather small. Therefore, to examine
the role of initial correlations for the reset in more detail, we artificially
add correlations to the factorizing initial state
\eqref{eq:model:state_fact_init},
\begin{align} \label{eq:model:state_add_corr}
  \op{\rho}^{\text{init}}_3
  =
  \op{\rho}_{\text{Q}}^{\text{th}} \otimes \op{\rho}_{\text{TLS}}^{\text{th}}
  +
  \begin{pmatrix}
    0 & 0 & 0 & 0 \\
    0 & 0 & \gamma & 0 \\
    0 & \gamma^{*} & 0 & 0 \\
    0 & 0 & 0 & 0
  \end{pmatrix}.
\end{align}
Motivated by Eq.~\eqref{eq:model:th_bipartite}, we chose $\gamma \in \mathbb{R}$
and $\gamma < 0$, while ensuring that the result is still a valid density
matrix.

We quantify the total amount of correlations in terms of the mutual information
$\mathcal{I}$ of qubit and memory TLS~\cite{Henderson2001}. This corresponds to
the total amount of correlations, both classical and quantum, between system and
environment since the qubit couples directly only to the TLS\@. To distinguish
between classical and quantum correlations, various concepts and measures haven
been introduced~\cite{Modi.RevModPhys.84.1655}. Here, we use quantum
discord~\cite{Ollivier2002} to quantify the amount of quantum correlations,
which is analytically computable for all considered
states~\cite{Ali.PRA.81.042105}. We also calculate entanglement in terms of
concurrence~\cite{Wootters1998}.

If not stated otherwise, $\omega_{\text{TLS}} \neq \omega_{\text{Q}}$ in the
following and in particular, $\omega_{\text{TLS}}>\omega_{\text{Q}}$. We set
$\hbar=1$ as well as $\omega_{\text{Q}}=1$ which define the units for time and
energy, respectively. The chosen parameters are typical for superconducting
qubits~\cite{DevoretSci2013}. In particular, our model could be easily
implemented by two superconducting qubits in an RLC circuit~\cite{Karimi2016},
where the resistor acts as a thermal reservoir, or by two superconducting qubits
with one of them coupled to a lossy cavity~\cite{Gerling2013}.

\section{Numerical Results} \label{sec:num}
The control problem of qubit reset with the equation of
motion~\eqref{eq:model:LvN} and initial
conditions~\eqref{eq:model:state_fact_init}, \eqref{eq:model:th_bipartite} and
\eqref{eq:model:state_add_corr} is not easily amenable to an analytical
solution. We therefore first determine optimized fields for the reset of the
qubit using numerical quantum optimal control~\cite{Glaser.EPJD.69.279}.

\subsection{Optimal Control Theory} \label{subsec:num:oct}
Assuming that a quantum system can be influenced by external fields
$\{\pulse_k(t)\}$, optimal control theory (OCT) provides the means to maximize
or minimize a predefined figure of merit. In our case, the control problem is
a simple state-to-state transfer~\cite{Bartana.JCP.106.1435}, achieved within
a fixed time $T$. The total optimization functional,
\begin{align}
  F\left[\left\{\pulse_{k}\right\}\right]
  =
  \error_{T}\left[\op{\rho}(T)\right]
  + \int_{0}^{T} \text{d}t\,
    g\left[\left\{\pulse_{k}(t)\right\}, \op{\rho}(t), t\right],
\end{align}
consists of the figure of merit $\error_{T}\left[\op{\rho}(T)\right]$ and
additional constraints, captured in a function $g$. In the following, we
consider only a single external field, $\pulse(t)$. Our figure of merit is the
error in preparing the qubit in the desired target state, irrespective of the
TLS state. This can be expressed as~\cite{Rojan.PRA.90.023824}
\begin{align} \label{eq:num:J_T}
  \error_{T}\left[\op{\rho}(T)\right]
  =
  1 - \left\langle
      \Psi^{\text{targ}}_{\text{Q}} \right|
      \tr{\text{TLS}}{}{\op{\rho}(T)}
      \left| \Psi^{\text{targ}}_{\text{Q}}
    \right\rangle\,,
\end{align}
where $\tr{\text{TLS}}{}{\cdot}$ describes the partial trace over the TLS\@.
Without loss of generality, we choose the target state
$\ket{\Psi^{\text{targ}}_{\text{Q}}}$ to be the bare ground state of the qubit.

We will use Krotov's method~\cite{Krotov.AutomRemContr.60.1427}, an iterative
optimization algorithm with built-in monotonic
convergence~\cite{Reich.JCP.136.104103}, in the following. The constraint
function is chosen as
\begin{align} \label{eq:num:g}
  g\left[\left\{\pulse(t)\right\}\right]
  = \frac{\lambda}{S(t)} \left(\pulse(t)
  - \pulse^{\text{ref}}(t)\right)^{2},
\end{align}
where $\lambda$ is a numerical parameter that controls the update magnitude of
the field $\pulse(t)$, $S(t)$ a shape function and $\pulse^{\text{ref}}(t)$
a reference field (taken to be the field from the previous iteration). The
actual update equation from the fields is determined by Eq.~\eqref{eq:num:g},
the equation of motion~\eqref{eq:model:LvN} and the final time
target~\eqref{eq:num:J_T}. For more details see Ref.
\cite{Reich.JCP.136.104103}.

\subsection{Factorizing Initial State} \label{subsec:num:fact}
We start by deriving the optimal reset protocol for a factorizing initial
state~\eqref{eq:model:state_fact_init} of qubit and TLS, i.e., when no initial
correlation between system and environment, i.e., between qubit and TLS, is
present. Note that the level splittings of qubit and TLS are not the same and
$\omega_{\text{TLS}} > \omega_{\text{Q}}$. This, together with the identical
temperature of qubit and TLS, results in a higher von Neumann entropy of qubit
than TLS\@. According to the second law of thermodynamics, one would expect the
best cooling to be achieved by an entropy exchange between TLS and qubit. This
has indeed been observed before~\citep{Horowitz2014,Pietro2016}.

\begin{figure*}[tb]
  \centering
  \includegraphics[width=0.99\linewidth]{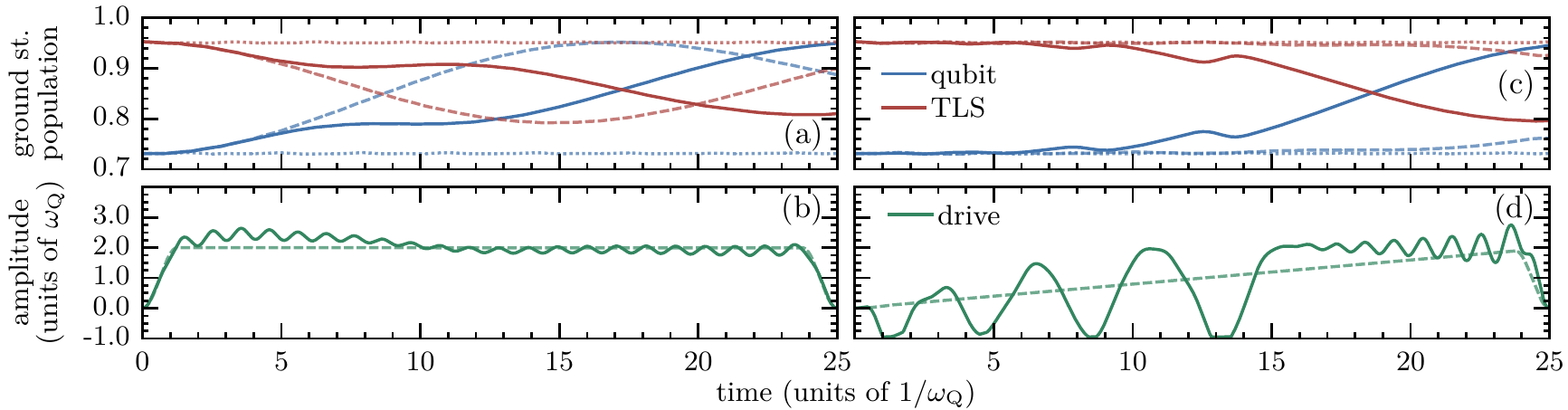}
  \caption{%
    (a,c) Population dynamics induced by the optimized fields (solid lines) for
    a factorizing initial state, Eq.~\eqref{eq:model:state_fact_init}. The
    corresponding fields are shown in (b,d). The dotted lines illustrate the
    free evolution of the system, the dashed lines the guess field and its
    evolution. Left and right hand side used different guess fields for the
    optimization. Parameters are $\omega_{\text{Q}}=1.0$,
    $\omega_{\text{TLS}}=3.0$, $J=0.1$, $\kappa=0.04$ and $\beta=1.0$. The
    initial ground state populations of qubit and TLS are
    $p_{\text{Q}}^{\text{init}} = 0.731$ and $p_{\text{TLS}}^{\text{init}}
    = 0.953$. The final value for the qubit's population is given by the
    fidelity $1 - \error_{T} = p_{\text{Q}}(T) \approx 0.950$ with error
    $\error_{T} = 5.04\%$ (a,b), respectively $\error_{T} = 5.44\%$ (c,d).
  }
  \label{fig:dyn_fact}
\end{figure*}
For the chosen parameters, entropy exchange can be realized by simply swapping
the ground state populations of qubit and TLS\@. This is best achieved when
qubit and TLS are in resonance. As can be seen in Eq.~\eqref{eq:model:hamQ}, the
control field $\pulse(t)$ effectively changes the frequency of the qubit.
Therefore an educated guess would be to ramp qubit and TLS rapidly into
resonance and stay there just long enough for a full swap operation.
Figure~\ref{fig:dyn_fact}(a) shows the dynamics for this particular guess field
(dashed lines), as well as the free evolution (dotted lines) and the dynamics
under the optimized field (solid lines). With the optimized field, we indeed
obtain the anticipated swap in the ground state populations at $t=T$. In
contrast, for the guess field, the maximal $p_{\text{Q}}(t)$ is already achieved
at $t\approx17$.

As we will show analytically in Sec.~\ref{sec:ana} below, the swap is the best
and fastest protocol for all factorizing initial conditions when the TLS is
initially diagonal in its eigenbasis. The analytical bounds for the minimal
error and the shortest possible duration in which the minimal error is reached,
given the parameters used in Fig.~\ref{fig:dyn_fact}, are
\begin{align} \label{eq:num:fact_limits}
  \error_{T}^{\text{min}}
  =
  1 - p_{\text{TLS}}^{\text{th}}
  =
  4.74\%,
  \quad
  T^{\text{min}}
  =
  \frac{\pi}{2 J}
  =
  15.7.
\end{align}
The actual value of the minimal error $\error_{T}^{\text{min}}$ is determined by
the initial ground state population $p_{\text{TLS}}^{\text{th}}$ of the TLS\@,
i.e., it is governed by the reservoir temperature. One might wonder why the
minimal time $t \approx 17$ required for the swap in Fig.~\ref{fig:dyn_fact}(a)
is larger than $T^{\text{min}}$ in Eq.~\eqref{eq:num:fact_limits}. This is due
to the fact that, for the sake of experimentally feasible control signals, we do
not allow $\pulse(t)$ to be instantaneously switched on and off. If we relax
this constraint, our optimized control reaches the quantum speed limit
$T^{\text{min}}$.

For any time longer than $T^{\text{min}}$, there is always at least one solution
achieving maximal cooling but there may be more,
 i.e., the control strategy is not unique. Another
possible control field is shown exemplarily in Fig.~\ref{fig:dyn_fact}(d). The
non-uniqueness of the solution allows for taking into account further
experimentally desirable features, such as restriction of the maximal amplitude
of the control, without losing performance.

\RED{%
One may wonder how robust these solutions are to noise in the controls or in the
initial state. We have quantified the robustness of the dynamics shown in
Fig.~\ref{fig:dyn_fact}(a) by averaging over $1000$ realizations of Gaussian
amplitude noise for the optimized field shown in Fig.~\ref{fig:dyn_fact}(b). For
a typical noise level of 1\% in the control amplitude, added in form of a varying scaling factor to the control, the final error increases by only a small amount, from 5.04\% to
5.16\% on average. In order to simulate noise in the initial state,
Eq.~\eqref{eq:model:state_fact_init}, we have added Gaussian noise to the input
parameters $\omega_{\text{Q}}$, $\omega_{\text{TLS}}$ and $\beta$, using again
$1000$ realizations. For noise levels up to 2\%, we obtain no change in the
error at all, and even 10\% of state noise  increase the error only from 5.04\%
to 5.18\% on average. The protocol is thus very robust with respect to noise in
the initial state. The reason for this finding will become clear below in
Sec.~\ref{sec:ana}.
}

\subsection{Correlated Initial State} \label{subsec:num:corr}
An obvious choice for a correlated initial state is the joint thermal
equilibrium state~\eqref{eq:model:th_bipartite} of qubit and TLS\@. For the
chosen parameters, the mutual information of this state is rather small,
$\mathcal{I}^{\text{init}} =4.0 \cdot 10^{-3}$. The state is separable but has
non-zero quantum discord. Note that all initial states studied within this
section have non-vanishing quantum discord, since for a thermalized TLS there
is no state with only classical correlations.

\begin{figure*}[tb]
  \centering
  \includegraphics[width=0.99\linewidth]{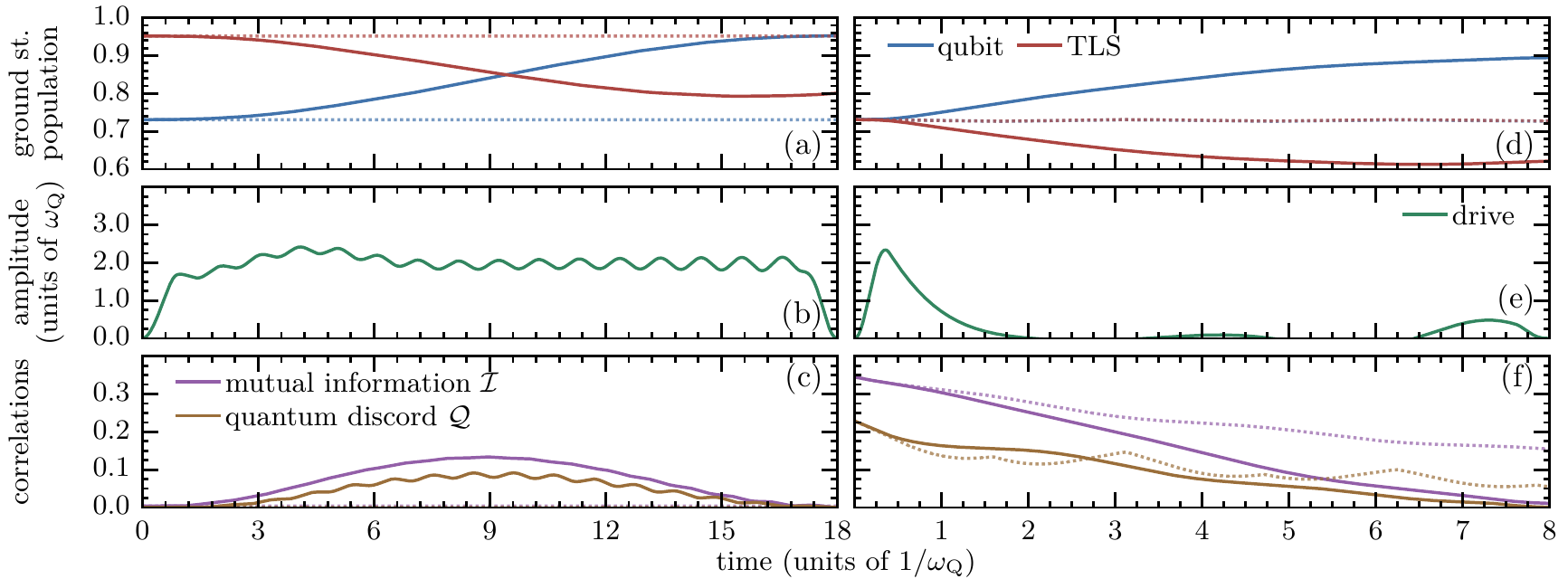}
  \caption{%
    Same as Fig.~\ref{fig:dyn_fact} but with correlated, non-entangled initial
    states. For the left hand side, the initial state is
    Eq.~\eqref{eq:model:th_bipartite} and, after optimization, the error at
    final time becomes $\error_{T} = 4.74\%$ and thus coincides with the limit
    $\error_{T}^{\text{min}}$, cf. Eq.~\eqref{eq:num:fact_limits}. For the right
    hand side, the initial state is
    Eq.~\eqref{eq:model:state_add_corr} with $\omega_{\text{Q}}
    = \omega_{\text{TLS}} = 1.0$ and $\gamma=-0.19$. With these level
    splittings, the error limit for factorizing thermal initial states amounts
    to $\error_{T}^{\text{min}} = 26.9\%$. It is given directly by the initial
    state since cooling is not possible at all in this case. With initial
    correlations, the error under the optimized field becomes
    $\error_{T}=10.52\%$ and is thus much smaller than $\error_{T}^{\text{min}}$
    for factorizing initial states.
  }
  \label{fig:dyn_corr}
\end{figure*}
As can be seen in Fig.~\ref{fig:dyn_corr}(a,b,c), both cooling and erasure of
correlation is achieved by the optimized control field. The final value of the
error in Fig.~\ref{fig:dyn_corr}(a), $\error_{T} = 4.74\%$, coincides with the
minimal error $\error_{T}^{\text{min}}$ for factorizing initial states, cf.
Eq.~\eqref{eq:num:fact_limits}. Optimal control therefore allows us to erase
initial correlations.
\RED{%
  A robustness analysis analogous to that for Fig.~\ref{fig:dyn_fact} yields
  very similar results: Amplitude noise at a level of 1\% increases the error
  from 4.74\% to 4.97\%, whereas noise in the state has no effect at all up to
  the 2\% level. It increases the error to only 4.85\% at the 10\% level.
}

To further investigate the role of initial correlations, we now choose qubit and
memory TLS to be in resonance, i.e. $\omega_{\text{Q}} = \omega_{\text{TLS}}$.
For factorizing initial conditions, no cooling at all would be possible.
Additionally, we enhance the correlations, the initial state is given by
Eq.~\eqref{eq:model:state_add_corr}. It is thermal in the sense that, if TLS or
qubit is traced out, one obtains Eq.~\eqref{eq:model:state_two_level_therm_eq}.
Surprisingly, we are not only able to erase the correlations, but even achieve
further cooling of the system, as can be seen in Fig.~\ref{fig:dyn_corr}(d,e,f),
for an initial state with mutual information $\mathcal{I}^{\text{init}} = 0.345$
and quantum discord $\mathcal{Q}^{\text{init}} = 0.228$. This is clear evidence
for system-environment correlations acting as a resource for cooling.

\begin{figure}[tb]
  \centering
  \includegraphics[width=0.99\linewidth]{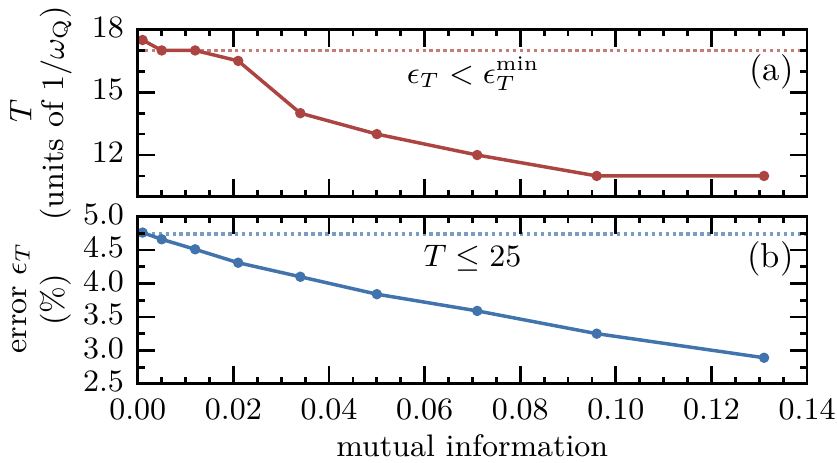}
  \caption{%
    Quantum speed limit (a) and minimal error (b) for a parametrical variation
    of the strength of initial correlations in
    Eq.~\eqref{eq:model:state_add_corr}. Note that upper and lower panel display
    results of different optimizations, only their initial states were
    identical.
    Panel (a) shows the smallest final time $T$, which still yields an error
    $\error_{T} < \error_{T}^{\text{min}}$. The dashed line corresponds to the
    approximate minimal time for a swap operation for factorizing initial
    states, taking into account finite ramps of the field at the beginning and
    end, cf. Fig.~\ref{fig:dyn_fact}(b).
    Panel (b) shows the smallest error $\error_{T}$ for any final time
    satisfying $T \leq 25$. The dashed line corresponds to the limit
    $\error_{T}^{\text{min}}$ for factorizing initial states, cf.
    Eq.~\eqref{eq:num:fact_limits}.
    Same parameters as in Fig.~\ref{fig:dyn_fact}.
  }
  \label{fig:corr_var}
\end{figure}
Remarkably, even the speed limit obtained for factorizing initial conditions
does not hold anymore. As can be seen in Fig.~\ref{fig:corr_var}(a), with
increasing total correlations, i.e., mutual information, the error threshold of
the factorizing dynamics, $\error_{T}^{\text{min}}$, can be reached in shorter
times. Note that although the upper left point in Fig.~\ref{fig:corr_var}(a)
lies above the approximate quantum speed limit for factorizing initial
conditions, this is only due to influence of the counter rotating terms (which
we will analyze in more detail in Appendix~\ref{app:RWA}). If we temporarily
neglect the counter rotating terms, the result coincides with the quantum speed
limit.

Moreover, Fig.~\ref{fig:corr_var}(b) shows that the final error $\error_{T}$ is
reduced for increasing initial correlations. While we have also studied
entangled initial states, the data is not presented here, as the results do not
differ. We find that only the amount of mutual information, i.e., the total
amount of correlations, not the type, i.e., classical or quantum correlations,
is relevant for cooling.

\RED{%
A natural question is whether the speed limit reported in
Fig.~\ref{fig:corr_var} depends on the type of control over the qubit. It turns
out that a control field that couples to the system via
$\op{\sigma}_{\mathrm{Q}}^{x}$ instead of $\op{\sigma}_{\mathrm{Q}}^{z}$ in
Eq.~\eqref{eq:model:hamQ} does not perform better (data not shown). We have
found solutions swapping the populations between qubit and memory TLS also for
that type of control when starting from factorizing initial states. Similarly to
$\op{\sigma}_{\mathrm{Q}}^{z}$-control, correlations in the initial state allow
for better reset with smaller errors. However, more time is required in both
cases when the control couples via $\op{\sigma}_{\mathrm{Q}}^{x}$. As
a consequence, the weakly coupled reservoir has a larger impact on the dynamics.
}

To summarize our findings obtained so far, it is not only possible to reset the
qubit in the presence of initial correlations; initial correlations between
system and environment can actually be used to enhance the performance of the
cooling protocol. Moreover, in the resonant case, initial correlations enable
cooling that is impossible without their presence. We analyze the dynamics that
lead to this surprising result in more detail in Sec.~\ref{sec:ana}.

\subsection{Non-Markovianity} \label{subsec:nonM}
Finally, we investigate whether non-Markovianity of the dynamics has any
influence on the optimized fields and achievable final errors. The dynamics of
the qubit becomes Markovian or non-Markovian depending on the ratio $J/\kappa$.
We quantify this by the accessible volume of state space~\cite{Lorenzo2013} to
study a possible interplay between non-Markovianity and control.

In our setup, we observe that non-Markovianity seems to be linked to population
flow between qubit and memory TLS\@. More precisely, a monotonic decrease in the
qubit's state space volume can be observed, when populations flows from the
memory TLS into the qubit, i.e., increasing the ground state population of the
qubit while decreasing it for the memory TLS\@. This hints towards Markovian
dynamics. In contrast, an increase in the state space volume occurs for the
reversed population flow, indicating non-Markovian dynamics.

The population flow between qubit and memory TLS is governed by their
\emph{effective} coupling. It is directly influenced by the coupling $J$ and
indirectly by the relative detuning $\delta(t) = \omega_{\text{Q}} + \pulse(t)
- \omega_{\text{TLS}}$ between both. The frequency, with which the population
flow changes its direction, increases with $|\delta(t)|$, while its smallest
value is assumed for $\delta(t)=0$, where the frequency is entirely determined
by $J$. According to this observation, the dynamics of the time-optimal solution
(cf. Eq.~\eqref{eq:num:fact_limits}) turns out to be Markovian. In this case,
the ground state population of the qubit is constantly increasing until reaching
its maximum at $T^{\text{min}}$.  For longer times and non-optimal driving, the
controlled dynamics can become non-Markovian, cf. Fig.~\ref{fig:dyn_fact}(c,d),
as the population flows in both directions at intermediate times. Nevertheless,
implementing a swap at $T > T^{\text{min}}$ is also possible with entirely
Markovian dynamics, cf. Fig.~\ref{fig:dyn_fact}(a,b). This shows that even
though non-Markovianity is not crucial for the qubit reset, it also is not
harmful in the sense that the optimization does not suppress non-Markovianity.

\section{Analytical Results} \label{sec:ana}
Two observations in the analysis of the numerical results presented above allow
us to simplify our model~\eqref{eq:model:LvN}: (i) Solutions obtained under the
RWA perform almost equally well in comparison with solutions when the
counter-rotating terms are taken into account (we discuss this in more detail in
Appendix~\ref{app:RWA}). In other words, although the RWA is not a good
approximation for the dynamics, it may be invoked to determine the controls.
(ii) Two different timescales are relevant to characterize the interaction of
the qubit with the environment---a fast one to dump the qubit's entropy into the
pseudo-mode, determined by the coupling $J$, and a slow one leading to
re-equilibration, determined by the coupling $\kappa$. Most importantly, the
re-equilibration dynamics will never increase the purity of qubit or TLS above
their steady state values. The minimum final error and time for the qubit reset
are therefore determined only by the fast timescale dynamics.

These observations suggest to neglect the dynamics associated with the slow
timescale and described by the Lindblad operators in Eq.~\eqref{eq:model:LvN} as
well as the counter-rotating terms in the Hamiltonian~\eqref{eq:model:hamQT}. As
a result, the reset control problem becomes amenable to an analytical
solution.

\subsection{Control Equations for Cooling a Qubit} \label{subsec:ana:geo_eq}
In the following we use concepts from geometric control theory
\cite{Jurdjevic.book}, where the idea consists in transforming the dynamical
equations of the system in such a way that the optimality condition can be
expressed analytically~\cite{Boscain.JMathPhys.43.2107, Sugny.PRA.76.023419}.
For ease of the derivation, we transform states and Hamiltonian into the
rotating frame. Neglecting the counter-rotating terms and the (slow)
equilibration with the reservoir, the equation of motion reads
\begin{subequations}
  \begin{align} \label{eq:ana:LvN_rotating}
    \im \frac{\text{d}}{\text{d}t} \op{\rho}'(t)
    &=
    \left[\op{H}'(t), \op{\rho}'(t)\right],
    \\
    \op{H}'(t)
    &=
    \begin{pmatrix}
      \frac{\text{d} \delta(t)}{\text{d} t} \frac{t}{2} & 0 & 0 & 0
      \\
      0 & \frac{\text{d} \delta(t)}{\text{d} t} \frac{t}{2} & J(t) e^{- \im
      \delta(t) t} & 0
      \\
      0 & J(t) e^{\im \delta(t) t} & - \frac{\text{d} \delta(t)}{\text{d} t}
      \frac{t}{2} & 0
      \\
      0 & 0 & 0 & - \frac{\text{d} \delta(t)}{\text{d} t} \frac{t}{2}
    \end{pmatrix},
  \end{align}
\end{subequations}
where $\delta(t) = \omega_{\text{Q}} + \pulse(t) - \omega_{\text{TLS}}$ is the
time-dependent detuning of qubit and TLS\@. For the sake of
generality, we account for a possible time-dependence $J=J(t)$ of the
coupling strength between qubit and TLS\@.

For the numerical optimization in section~\ref{sec:num}, the optimization target
was to reset the qubit in its ground state. Here, we choose a more general
approach and maximize the qubit's purity~\footnote{This is also possible in the
numerical optimization. However, the more complicated target functional requires
a significantly more sophisticated optimization
algorithm~\cite{Reich.JCP.136.104103}.}. The key idea in the following is to
chose a representation of the state $\op{\rho}'(t)$ in terms of a set of real
variables $\left\{x_{1}(t), \dots, x_{16}(t)\right\}$ to span the entire state
space of qubit and TLS\@. Inserting this representation into
Eq.~\eqref{eq:ana:LvN_rotating}, one obtains coupled equations for all $x_{i}$.
In order to decouple these equations and reduce the number of relevant
variables, one needs to perform an appropriate variable transformation
$\left\{x_{1}(t), \dots, x_{16}(t)\right\} \rightarrow \left\{z_{1}(t), \dots,
z_{16}(t)\right\}$. A more detailed description of the transformations can be
found in Appendix~\ref{app:derivation}.

In the new variables, the qubit's purity becomes
\begin{align} \label{eq:ana:qubit_purity_z}
  \mathcal{P}_{\text{Q}}
  =
  \frac{1}{2} + 2 \left(%
    z_{1}^{2} + z_{5}^{2} + z_{7}^{2}
  \right)\,,
\end{align}
where we have dropped the explicit time dependence for all quantities. The
corresponding equations of motion are decoupled into two separate subspaces. On
the one hand, we have
\begin{align} \label{eq:ana:veceqs_S1}
  \begin{pmatrix}
    \dot{z}_{1} \\ \dot{z}_{2} \\ \dot{z}_{3}
  \end{pmatrix}
  =
  2 J_{1}
  \begin{pmatrix}
    - z_{2} \\ z_{1} - z_{1}^{\text{c}} \\ 0
  \end{pmatrix}
  +
  2 J_{2}
  \begin{pmatrix}
    - z_{3} \\ 0 \\ z_{1} - z_{1}^{\text{c}}
  \end{pmatrix}
  +
  2 \alpha
  \begin{pmatrix}
    0 \\ - z_{3} \\ z_{2}
  \end{pmatrix},
\end{align}
describing the dynamics of the qubit's ground state population, $p_{\text{Q}}
= z_{1} + 1/2$, within the three-dimensional subspace $S_{1} = \left\{z_{1},
z_{2}, z_{3}\right\}$, $z_{1}^{\text{c}}$ being a constant.
Note that $z_{2}$, $z_{3}$ are non-zero at time $t=0$ only if initial
correlations are present, cf. Eqs.~\eqref{eq:model:state_add_corr},
\eqref{eq:ana:rho} and~\eqref{eq:app:z_transform}.
Equation~\eqref{eq:ana:veceqs_S1} thus already indicates that initial
correlations can be transferred into ground state population and hence purity.
On the other hand, the qubit's coherences, $\gamma_{\text{Q}}= z_{5} + \im
z_{7}$, evolve within the four-dimensional subspace $S_{2}=
\left\{z_{5},z_{6},z_{7},z_{8}\right\}$,
\begin{align} \label{eq:ana:veceqs_S2}
  \begin{pmatrix}
    \dot{z}_{5} \\ \dot{z}_{6} \\ \dot{z}_{7} \\ \dot{z}_{8}
  \end{pmatrix}
  =
  J_{1}
  \begin{pmatrix}
    - z_{6} \\ z_{5} \\ z_{8} \\ - z_{7}
  \end{pmatrix}
  +
  J_{2}
  \begin{pmatrix}
    z_{8} \\ - z_{7} \\ z_{6} \\ - z_{5}
  \end{pmatrix}
  +
  2 \alpha
  \begin{pmatrix}
    z_{7} \\ 0 \\ - z_{5} \\ 0
  \end{pmatrix},
\end{align}
where $z_6$ and $z_8$ are related to the TLS coherence. The three fields are
given by
\begin{align}\label{eq:alpha}
  J_{1} = J \cos(\delta t),
  \quad
  J_{2} = J \sin(\delta t),
  \quad
  \alpha = \frac{1}{2} \frac{\text{d} \delta}{\text{d} t} t.
\end{align}

It is straightforward to show that the dynamics within the subspaces $S_{1}$ and
$S_{2}$ is restricted to the surface of two spheres. For $S_{1}$, we find from
Eq.~\eqref{eq:ana:veceqs_S1}
\begin{align} \label{eq:ana:spherical_eq_S1}
  \frac{\text{d}}{\text{d}t} R_{1}^{2}
  =
  0,
  \qquad
  R_{1}
  =
  \sqrt{\left(z_{1}-z_{1}^{\text{c}}\right)^{2} + z_{2}^{2} + z_{3}^{2}},
\end{align}
with $R_{1}$ the radius of the sphere centered around
$\left(z_{1}^{\text{c}}, 0, 0\right)$ with constant $z_{1}^{\text{c}}
= -(z_{4}+1)/2$, cf. Eq.~\eqref{eq:app:z_transform}. Similarly for $S_{2}$,
Eq.~\eqref{eq:ana:veceqs_S2} yields
\begin{align}
  \frac{\text{d}}{\text{d}t} R_{2}^{2}
  =
  0,
  \qquad
  R_{2}
  =
  \sqrt{z_{5}^{2} + z_{6}^{2} + z_{7}^{2} + z_{8}^{2}},
\end{align}
with radius $R_{2}$ and center $\left(0, 0, 0, 0\right)$. The values of $R_{1}$
and $R_{2}$ are determined by the initial values $z_{i}^{\text{init}}$ with
$i=1,\dots,8$. In other words, the accessible part of the entire state space is
fully determined by the initial state $\op{\rho}^{\text{init}} = \op{\rho}(0)
= \op{\rho}'(0)$.

\subsection{Optimal Strategy for Thermal Factorizing Initial States}
\label{subsec:ana:geo_fact_therm}
The factorizing initial state~\eqref{eq:model:state_fact_init} is obviously
diagonal. Thus we have $z_{2}^{\text{init}} = z_{3}^{\text{init}} = 0$ as well
as $z_{i}^{\text{init}} = 0$, $i=5, \dots, 8$. As a consequence, $R_{2} = 0$,
i.e., no dynamics will occur in $S_{2}$, and the relevant subspace is entirely
given by $S_{1}$. In the following, we parametrize
Eq.~\eqref{eq:model:state_fact_init} as
\begin{align} \label{eq:ana:init_state_q_times_t}
  \op{\rho}^{\text{init}}
  =
  \op{\rho}_{\text{Q}}^{\text{th}} \otimes
  \op{\rho}_{\text{TLS}}^{\text{th}}
  =
  \begin{pmatrix}
    a_{\text{Q}} & 0 \\ 0 & b_{\text{Q}}
  \end{pmatrix}
  \otimes
  \begin{pmatrix}
    a_{\text{TLS}} & 0 \\ 0 & b_{\text{TLS}}
  \end{pmatrix},
\end{align}
and assume $\op{\rho}_{\text{TLS}}^{\text{th}}$ to be initially more pure than
$\op{\rho}_{\text{Q}}^{\text{th}}$. This amounts to $a_{\text{TLS}}^{2}
+ b_{\text{TLS}}^{2} > a_{\text{Q}}^{2} + b_{\text{Q}}^{2}$ with $a,b$ the
ground and excited state populations of qubit and TLS, respectively. We first
discuss the resonant case, i.e., $\delta=0$ for all $t$, and derive the
time-optimal solution for the control problem. Second, we show that allowing for
$\delta \neq 0$ does not improve the best possible final purity of the qubit.

\begin{figure}[t]
  \centering
  \includegraphics[width=0.99\linewidth]{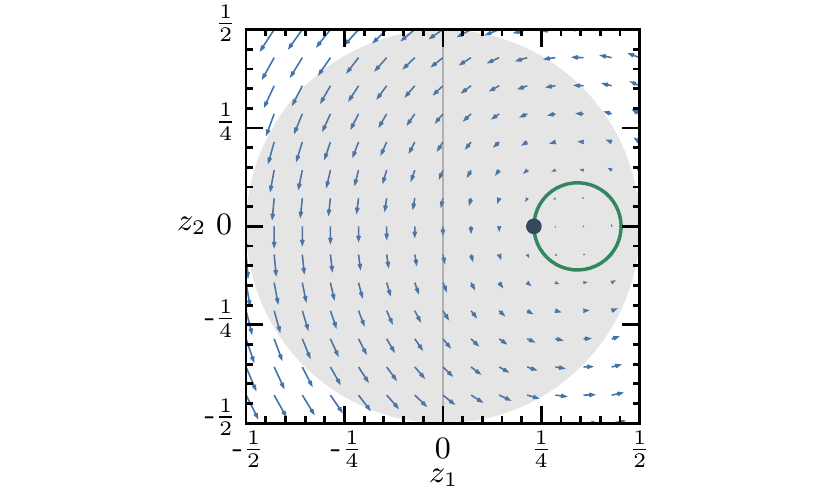}
  \caption{%
    Evolution of the qubit ground state population $p_{\text{Q}} = z_{1} + 1/2$
    (green line) within the subspace $S_{1}^{2}$ for non-vanishing coupling
    strength $J\neq0$ and factorizing initial state
    \eqref{eq:model:state_fact_init}, indicated by the large dot (parameters as
    in Fig.~\ref{fig:dyn_fact}). Qubit and TLS are in resonance ($\delta=0$ for
    all $t$) and the evolution of the state along the green line is determined
    by the vector field~\eqref{eq:ana:veceqs_S1_2} (blue arrows). The gray
    vertical line indicates the minimal purity (respectively, ground state
    population) of the qubit, cf. Eq.~\eqref{eq:ana:qubit_purity_z} with $z_{5}
    = z_{7}= 0$. The gray sphere in the background visualizes the projection of
    the entire state space onto the two-dimensional subspace $\left\{z_{1},
    z_{2}\right\}$.
  }
  \label{fig:ana:gct_therm_sep}
\end{figure}
For $\delta=0$ for all $t$, which implies $J_{1} = J$ and $J_{2} = \alpha = 0$,
Eq.~\eqref{eq:ana:veceqs_S1} is further simplified and the dynamics are confined
to the two-dimensional subspace $S_{1}^{2} = \left\{z_{1},
z_{2}\right\}$,
\begin{align} \label{eq:ana:veceqs_S1_2}
  \begin{pmatrix}
    \dot{z}_{1} \\ \dot{z}_{2}
  \end{pmatrix}
  =
  2 J
  \begin{pmatrix}
    - z_{2} \\ z_{1} - z_{1}^{\text{c}}
  \end{pmatrix}.
\end{align}
Figure~\ref{fig:ana:gct_therm_sep} shows the accessible state space for the
dynamics within $S_{1}^{2}$ when starting in the initial state used in
Fig.~\ref{fig:dyn_fact}. Depending on the sign of $J$, the initial state evolves
along the vector field ($J>0$) or opposite to it ($J<0$), cf.
Eq.~\eqref{eq:ana:veceqs_S1_2}. The optimization target can then be trivially
identified as the point with maximal $z_{1}$ on this curve. Assuming constant
positive coupling $J$, the state will evolve with constant speed along the green
line in Fig.~\ref{fig:ana:gct_therm_sep}. It then takes $T^{\text{min}} = \pi/(2
J)$ to reach the rightmost point. This can simply be shown by integrating along
the green line. Allowing for time-dependent coupling $J(t) \geq 0$, the minimal
time is given by
\begin{align} \label{eq:ana:T_opt}
  \int_{0}^{T^{\text{min}}} J(t) \text{d}t
  =
  \frac{\pi}{2}.
\end{align}
Therefore, the time-optimal solution is to choose $J(t)$ maximal for all $t$.

The point of maximum qubit purity, $\mathcal{P}_{\text{Q}}^{\text{max}}$, is
determined by the center $z_{1}^{\text{c}}$ of the sphere and its radius
$R_{1}$,
\begin{align} \label{eq:ana:P_q_max}
  \mathcal{P}_{\text{Q}}^{\text{max}}
  =
  \frac{1}{2} + 2 \left(z_{1}^{\text{c}} + R_{1}\right)^{2}
  =
  a_{\text{TLS}}^{2} + b_{\text{TLS}}^{2}
  =
  \mathcal{P}_{\text{TLS}}^{\text{init}},
\end{align}
with $\mathcal{P}_{\text{TLS}}^{\text{init}}$ the initial TLS purity.
Equations~\eqref{eq:ana:T_opt} and \eqref{eq:ana:P_q_max} hold for any initial
factorizing state of the form~\eqref{eq:ana:init_state_q_times_t} with the TLS
initially purer than the qubit. Note that for $z_{1}^{\text{c}} < 0$,
Eq.~\eqref{eq:ana:P_q_max} becomes $\mathcal{P}_{\text{Q}}^{\text{max}}
= \frac{1}{2} + 2 \left(z_{1}^{\text{c}} - R_{1}\right)^{2}$ but yields
identical results.

It is straightforward to see that a non-vanishing time-dependent detuning
$\delta \neq 0$ does not provide access to states with higher qubit purity. The
dynamics is confined to the surface of the three-dimensional sphere $S_{1}$, cf.
Eq.~\eqref{eq:ana:spherical_eq_S1}, and the point of maximal purity is already
accessible with $\delta=0$ for all $t$. It is important to note that $\delta
\neq 0$ involves dynamics in the $z_{3}$-dimension. This becomes crucial when
starting with initially correlated states.

\subsection{Optimal Strategy for Factorizing Initial States with
Coherences}
\label{subsec:ana:geo_fact_coh}
\begin{figure*}[t]
  \centering
  \includegraphics[width=0.99\linewidth]{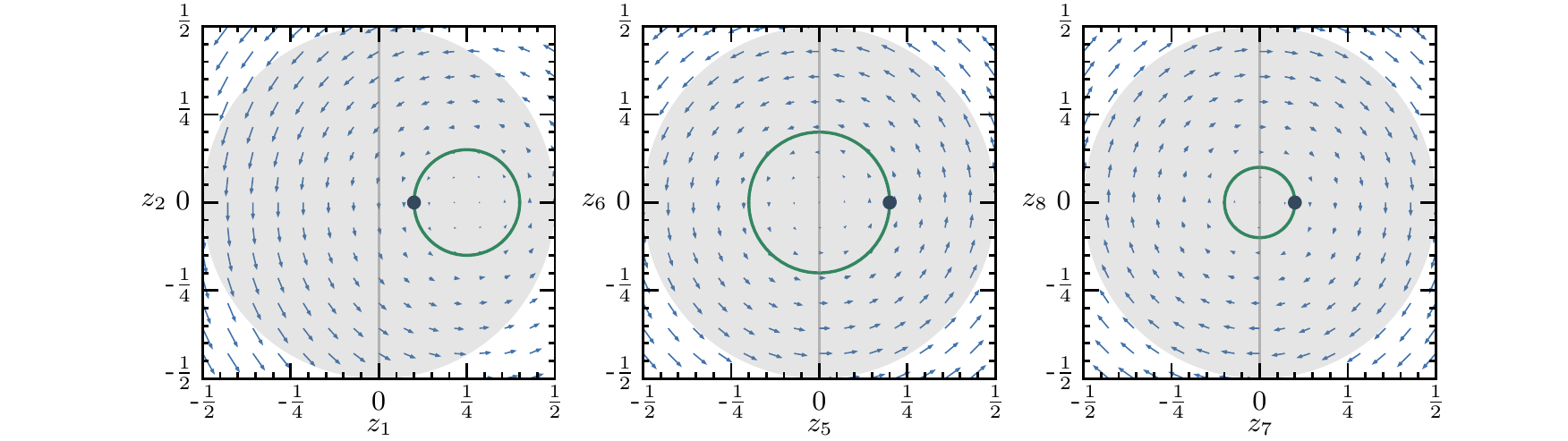}
  \caption{%
    Time evolution (green lines) within the three subspaces $S_{1}^{2}$,
    $S_{2}^{2}$ and $S_{3}^{2}$ (from left to right) for a factorizing initial
    state~\eqref{eq:ana:init_state_q_times_t_coh} with $a_{\text{Q}} = 0.6$,
    $b_{\text{Q}} = 0.4$, $\gamma_{\text{Q}} = 0.2 + \im 0.1$ and
    $a_{\text{TLS}} = 0.9$, $b_{\text{TLS}} = 0.1$, $\gamma_{\text{TLS}} = 0$.
    Qubit and TLS are in resonance ($\delta=0$ for all $t$). The dots indicate
    the initial state within the specific subspace, which then evolves along the
    vector fields~\eqref{eq:ana:veceqs_S1_2} and
    \eqref{eq:ana:veceqs_S2_2_and_S3_2}, represented by the blue arrows. The
    gray vertical lines indicate the respective minimal contribution to the
    qubit's purity for each subspace, while the gray spheres visualize the
    projection of the entire state space onto the subspaces.
  }
  \label{fig:ana:gct_Qcoh}
\end{figure*}
The most general initially factorizing state for qubit and TLS is given by
\begin{align} \label{eq:ana:init_state_q_times_t_coh}
  \op{\rho}^{\text{init}}
  =
  \op{\rho}_{\text{Q}}\otimes
  \op{\rho}_{\text{TLS}}
  =
  \begin{pmatrix}
    a_{\text{Q}} & \gamma_{\text{Q}} \\ \gamma_{\text{Q}}^{*} & b_{\text{Q}}
  \end{pmatrix}
  \otimes
  \begin{pmatrix}
    a_{\text{TLS}} & \gamma_{\text{TLS}} \\ \gamma_{\text{TLS}}^{*}
    & b_{\text{TLS}}
  \end{pmatrix},
\end{align}
with $a, b$ as in Eq.~\eqref{eq:ana:init_state_q_times_t} and
$\gamma_{\text{Q}}, \gamma_{\text{TLS}}$ the coherences of qubit and TLS\@. We
first consider the case $\gamma_{\text{TLS}} = 0$. From a physical perspective,
this is a well justified initial state, since we assume the TLS to be in
permanent contact with the reservoir and thus in thermal equilibrium. In
contrast, for the qubit, non-zero coherences, $\gamma_{\text{Q}} \neq 0$, are
a possible scenario, e.g., as a result of its previous use in a computation. In
this case, we again find $z_{2}^{\text{init}} = z_{3}^{\text{init}} = 0$.
However, $z_{5} = \rp\left\{\gamma_{\text{Q}}\right\}$ or $z_{7}=
\ip\left\{\gamma_{\text{Q}}\right\}$ or both will be non-zero. Note that
$z_{6}^{\text{init}} = z_{8}^{\text{init}} = 0$ still holds but there is
dynamics within the subspace $S_{2}$, since $R_{2} \neq 0$.

Assuming resonance in the following (i.e., $\delta=0$ for all $t$), the dynamics
within $S_{1}$ is reduced to the two-dimensional subspace $S_{1}^{2}$, as
discussed before. Similarly, the dynamics in the four-dimensional subspace
$S_{2}$ decouple and can be described by two two-dimensional subspaces,
$S_{2}^{2}$ and $S_{3}^{2}$. Their respective equations of motions are
\begin{align} \label{eq:ana:veceqs_S2_2_and_S3_2}
  \begin{pmatrix}
    \dot{z}_{5} \\ \dot{z}_{6}
  \end{pmatrix}
  =
  J
  \begin{pmatrix}
    - z_{6} \\ z_{5}
  \end{pmatrix},
  \qquad
  \begin{pmatrix}
    \dot{z}_{7} \\ \dot{z}_{8}
  \end{pmatrix}
  =
  J
  \begin{pmatrix}
    z_{8} \\ - z_{7}
  \end{pmatrix}.
\end{align}

Figure~\ref{fig:ana:gct_Qcoh} shows the evolution in the three subspaces
$S_{1}^{2}$, $S_{2}^{2}$ and $S_{3}^{2}$ for an exemplary initial factorizing
state with $\gamma_{\text{Q}} \neq 0$ and $\gamma_{\text{TLS}} = 0$. We now have
dynamics in all three subspaces. As before, maximizing the qubit's ground state
population, $p_{\text{Q}} = z_{1} + 1/2$, requires the time $T=\pi/(2J)$, which
corresponds to evolution in terms of a half circle in $S_{1}^{2}$. Importantly,
the motion within $S_{1}^{2}$ is twice as fast as that in $S_{2}^{2}$ and
$S_{3}^{2}$, which can be easily seen by comparing
Eqs.~\eqref{eq:ana:veceqs_S1_2} and \eqref{eq:ana:veceqs_S2_2_and_S3_2}.
Therefore, at time $T=\pi/(2J)$, the qubit's coherences, $\gamma_{\text{Q}}
= z_{5} + \im z_{7}$, vanish, since the evolution within $S_{2}^{2}$ and
$S_{3}^{2}$ only runs through a quarter circle. The minimal reset time is thus
not changed when allowing for coherences in the initial qubit state. This
finding is in line with the observation that for pure states (as considered in
this section), standard quantum speed limit bounds coincide with the bound
obtained from the Wigner-Yanase skew information which particularly quantifies
the coherence of a state (relative to the eigenbasis of the
Hamiltonian)~\footnote{For mixed states, these bounds do not coincide, and the
Wigner-Yanase skew information provides a tighter bound, highlighting the role
of coherences for the speed of evolution~\cite{Marvian.PRA.93.052331,
Pires.PRX.6.021031}.} \cite{Marvian.PRA.93.052331, Pires.PRX.6.021031}.
Moreover, as long as the initial purities of qubit and TLS satisfy
$\mathcal{P}_{\text{Q}}^{\text{init}} < \mathcal{P}_{\text{TLS}}^{\text{init}}$,
the time-optimal solution is still the swap operation given by Eqs.
\eqref{eq:ana:T_opt} and~\eqref{eq:ana:P_q_max}. This is true irrespective of
the specific initial state of the qubit.

If we allow for coherences also in the initial state of the TLS,
$\gamma_{\text{TLS}} \neq 0$, this does not hold anymore. In this case, some or
all of the initial values $z_{2}^{\text{init}}$, $z_{3}^{\text{init}}$,
$z_{6}^{\text{init}}$ and $z_{8}^{\text{init}}$ are non-zero. Geometrically, the
large dots in the three spheres $S_{1}^{2}$, $S_{2}^{2}$ and $S_{3}^{2}$ in
Fig.~\ref{fig:ana:gct_Qcoh} are then placed at arbitrary points along the green
curves. Thus, the evolution cannot easily be synchronized in terms of half and
quarter circles. Rather, exact knowledge of the initial state would be required
to determine the optimal solution.

\subsection{Optimal Strategy for Correlated Initial States}
\label{subsec:ana:geo_corr}
For correlated initial states, the dynamics involving the qubit ground state
population $z_1$ explores all three dimensions of the subspace $S_1$ spanned by
$z_{1}, z_{2}, z_{3}$. We show that a geometric analysis is still useful in this
case since it provides physical insight into the control mechanisms of the
optimal solution. In particular, it explains why initial correlations result in
a higher purity and a shorter time for the reset.

For any initial state satisfying Eq.~\eqref{eq:model:state_add_corr}, no
dynamics occurs in $S_{2}^{2}$ and $S_{3}^{2}$. It is then straightforward to
show that these correlated initial states allow to access states with higher
purity than factorizing states: Since the reduced states of qubit and TLS are
unchanged by the presence of correlations, the center $\left(z_{1}^{\text{c}},
0, 0\right)$ of the sphere in $S_{1}$ remains the same, while its radius $R_{1}$
increases, cf. Eq. \eqref{eq:ana:spherical_eq_S1}. As a result, the set of
accessible states that may be reached by the dynamics is enlarged.

\begin{figure}[t]
  \centering
  \includegraphics[width=0.99\linewidth]{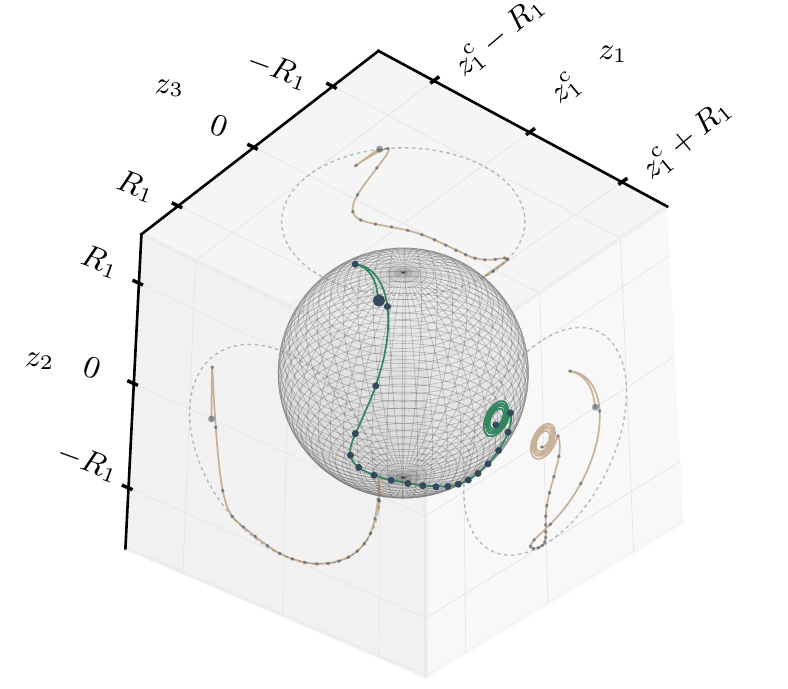}

  \includegraphics[width=0.99\linewidth]{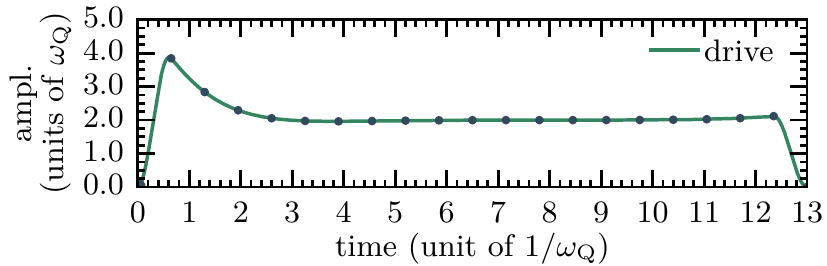}
  \caption{%
    Evolution within the subspace $S_{1}$ (top panel) for a correlated initial
    state of the form~\eqref{eq:model:state_add_corr} under the optimized field
    shown in the bottom panel ($\gamma=-0.09$, all other parameters as in
    Fig.~\ref{fig:dyn_fact}). The large dot marks the initial point in state
    space, the small dots indicate the evolution in chunks of $5\%$ of the total
    time. The final error is $\error_{T} = 1.6\%$, achieved within $T = 13$.
  }
  \label{fig:ana:gct_corr}
\end{figure}

Figure~\ref{fig:ana:gct_corr} shows the evolution starting from a correlated
initial state under a field designed by numerical optimization. It illustrates
why the quantum speed limit for factorizing initial states can be beaten. For
the initial state in Fig.~\ref{fig:ana:gct_corr}, $z_{2}^{\text{init}} = 0$ and
$z_{3}^{\text{init}} < 0$. The optimized field drives the state rapidly towards
the $z_{3}=0$ plane. This is achieved by the characteristic off-resonant peak in
the optimized field between $t=0$ and $t=2$. The subsequent evolution with
$\delta=0$ becomes two-dimensional within the $z_{1}$-$z_{2}$ plane; it is
equivalent to that in Fig.~\ref{fig:ana:gct_therm_sep} discussed above. However,
in contrast to the dynamics shown in Fig.~\ref{fig:ana:gct_therm_sep}, the
motion in the $z_{1}$-$z_{2}$ plane has to overcome a reduced distance as
a consequence of the initial transfer between $z_{3} < 0$ and $z_{3} = 0$. It
can be seen from the projection of the entire motion onto the $z_{1}$-$z_{2}$
plane (shown in the front left plane in Fig.~\ref{fig:ana:gct_corr} top, note in
particular the position of the third small dot), that less than a half circle
has to be overcome by the evolution with $\delta=0$ to reach the point of
largest purity, $z_{1,\text{max}}=z_1^{\text{c}}+R_1$. Since the initial
transfer towards the $z_3=0$ plane is accomplished faster than any motion within
this plane, the total time is reduced. Unfortunately, however, the reduction in
time comes at a cost, namely the control field must be tuned to the initial
value of $z_3$. In other words, for correlated initial states, derivation of the
optimal control strategy requires knowledge of the initial state.

This analysis can be completed by a geometric description of the solution. To
this end, we consider the differential system \eqref{eq:ana:veceqs_S1} and
assume the coupling $J$ to be bounded, while there is no constraint on
$\frac{\text{d}\delta}{\text{d}t}$, i.e., on $\alpha(t)$, cf.
Eq.~\eqref{eq:alpha}. As in the numerical optimization, the optimal solution can
be decomposed into two steps. In a first stage, we neglect the first two terms
on the right hand side of Eq.~\eqref{eq:ana:veceqs_S1} and the $\alpha$-term is
used to move arbitrarily fast in the $z_2$-$z_3$ plane from the initial point
into the $z_3=0$ plane. This motion is completed in a short time
$\tau_\varepsilon$ provided $\alpha(t)$ satisfies the condition
\begin{align}
  \int_0^{\tau_\varepsilon} 2\alpha(t) \text{d}t
  =
  \frac{\pi}{2},
\end{align}
which, after integration by parts, leads to
\begin{align}
  \delta(\tau_\varepsilon) - \int_0^{\tau_\varepsilon}\delta(t) \text{d}t
  =
  \frac{\pi}{2}.
\end{align}
A standard solution for $\delta$ is given by a linear time evolution of the form
\begin{align}
  \delta(t)
  =
  \frac{\pi t}{2\tau_\varepsilon(1-\tau_\varepsilon/2)}
  \quad \mathrm{for} \quad
  t \in [0,\tau_\varepsilon].
\end{align}
The second part of the optimal solution is the meridian trajectory in the
$z_3=0$ plane with $\delta(t)=0$. In fact, for $\delta(t)$ constant, we recover
the Grushin model~\cite{Sugny.PRA.77.063420}. It can be shown (using the
appendix of Ref.~\cite{Sugny.PRA.77.063420}) that the meridian trajectory is
the solution minimizing the time to reach the state of largest purity,
$z_{1,\textrm{max}}$. The time required for the motion along the meridian is
fixed by the initial point of this dynamics, it is
$T^{\textrm{min}}=\theta^{\text{init}}/(2J)$ where $\theta^{\text{init}}$ is
the polar angle of the sphere $S_1$ given by $z_{1}^{\text{init}} = R_1\cos
(\theta^{\text{init}})$. Assuming the time to reach the $z_3=0$ plane,
$\tau_\varepsilon$, to be arbitrarily small, the time
$\tau_\varepsilon+T^{\textrm{min}}$ required for both steps of the time-optimal
solution for correlated initial states is smaller than the time of $\pi/(2J)$
obtained with factorizing initial states. This rigorously confirms the role of
initial correlations for the speedup of the purification process.

\RED{The robustness of the numerical control solutions with respect to noise in either control amplitude or initial state, observed in Sec.~\ref{sec:num}, can be rationalized by the analytical solutions found here. Key to all of the reset strategies is a population swap between qubit and TLS. This is independent of the actual populations, as evidenced in the remarkable robustness with respect to noise in the initial state. The population swap  
requires resonance between qubit and TLS. Amplitude noise up to a level of 1\% does not perturb the resonance sufficiently to have a noticeable effect on the final errors.
}

\section{Summary and Conclusions} \label{sec:conclusions}
We have shown that quantum optimal control theory allows to derive protocols for
qubit reset with minimal error in minimum time. Such fast and reliable qubit
reset is crucial for quantum devices to be used multiple times or quantum
machines to operate in a cyclic way. Our main assumption was that the qubit is
coupled to a structured environment, consisting of a pseudo-mode and
a reservoir. Note that introducing more than one pseudo-mode will not change the
overall picture since the reset will be determined by the most strongly coupled
mode, in analogy with Ref.~\cite{Reich2015}. The coupling to the pseudo-mode is
taken to be small compared to the level spacings but large enough to render the
qubit dynamics non-Markovian; the coupling to the reservoir is weak.
\RED{We have assumed the system-pseudomode coupling to be of
$\op{\sigma}_x\op{\sigma}_x$-type. This is motivated by the fact that cooling
requires population exchange.}
In an actual experiment, the pseudo-mode could be realized by an ancilla, and
the reservoir by a resistor or a lossy cavity---scenarios that are found for
example in superconducting circuits.

The assumptions of our model imply two timescales---a fast one for the
interaction between qubit and TLS (pseudo-mode) and a slow one for
re-equilibration with the reservoir. This timescale separation allows to solve
the reset control problem analytically and evaluate the bounds for minimum error
and minimum time for certain initial states and under the rotating wave
approximation. Assuming the TLS to be initially in thermal equilibrium with the
reservoir, we find different solutions to the control problem for factorizing
and correlated initial states. If qubit and TLS are initially uncorrelated (and
thus there are no correlations between qubit and all of the environment), the
time-optimal solution is a swap operation. Cooling and reset are thus only
possible if the TLS is initially colder, i.e., purer, than the qubit. The
minimal error is determined by the temperature as well as the initial difference
in the qubit and TLS level splittings, it becomes smaller for larger TLS
splitting. The minimal time is set by the coupling strength between qubit and
TLS\@. The time-optimal solution consists in ramping qubit and TLS into and out
off resonance.
\RED{Since this is most easily achieved by an external control field coupling to
the system via $\op{\sigma}_z$, $\op{\sigma}_z$-controls outperform controls
coupling to the system via $\op{\sigma}_x$. The time-optimal solution}
is valid for all factorizing initial states of qubit and TLS (with the TLS
initially in thermal equilibrium with the reservoir), i.e., no a priori
knowledge of the initial qubit state is necessary.
If initial correlations between qubit and environment are present, the limits on
minimum error and minimum time for the uncorrelated case both can be beaten.
However, in this case, knowledge of the initial state is required to derive the
reset protocol since the control strategy is tied to the amount of initial
correlations. This information is easily accessible, if the initial state is
e.g.\ the steady state of a non-weakly coupled system.

\RED{%
The control technique that we have employed here is open loop which is the
method of choice when one seeks time-optimal
solutions~\cite{Glaser.EPJD.69.279}. There also exist a number of closed-loop
feedback control approaches to qubit purification. They are based on continuous
measurement and use feedback to control the qubit in such a way that the qubit's
purification rate increases~\cite{Combes.PRL.96.010504, Wiseman.NJP.8.90,
Wiseman.QIP.7.71, Combes.PRA.82.022307}.
While the requirement of carrying out measurements is the price to pay with
closed-loop approaches, they come with the advantage of inherent robustness to
noise. In contrast, open-loop control per se is not robust to noise, although it
can be made so~\cite{Rojan.PRA.90.023824,GoerzPRA14}.  We have therefore
assessed the robustness of our control solutions by adding Gaussian-distributed
noise to both the amplitude of the control and to the initial state.  Our
solutions are robust to amplitude noise up to about 1\%. When realizing our
model consisting of a qubit and a pseudo-mode with two superconducting qubits,
such a noise level by far exceeds typical experimental
values~\cite{Quintana.PRL.118.057702}. Moreover, we have found noise in the
initial state to not affect the final reset error all the way up to a level of
10\%. This remarkable robustness is explained by the time-optimal control
strategy consisting in a population swap between qubit and pseudo-mode.
}

Both speed-up and error reduction in the presence of initial correlations can be
understood by the geometry of the evolution in state space.  Remarkably, even in
the case where qubit and TLS are initially in resonance and cooling would not be
possible at all for factorizing initial conditions, correlations allow for
entropy export. Initial correlations with the environment thus act as a resource
for the qubit reset. Quantifying the initial correlations in terms of the mutual
information, quantum discord and entanglement of qubit and TLS, we have found
the amounts by which error and time can be reduced to be directly linked to the
mutual information. In contrast, the type of correlation turns out not to play
any role. In other words, entanglement between system and environment is not
required and classical, or at least quantum correlation without entanglement,
are sufficient to beat the limits on error and time for factorizing initial
states.

Our findings suggest to actively exploit initial correlations between qubit and
environment in qubit reset, using either a single ancilla qubit or true defect.
For example for superconducting qubits, the latter can be characterized
precisely both in terms of level splitting and
coupling~\cite{Shalibo.PRL.105.177001} and thus effectively act like an
ancilla~\cite{Reich2015}. For optimum performance of the qubit reset, the amount
of initial correlations must be known. The idea is then to engineer the initial
correlations between the qubit and its environment before carrying out the
reset. This is related to algorithmic cooling where correlations are created
dynamically by cross-relaxation~\cite{Rodriguez2017a} or measurements of
interacting qubits~\cite{Rodriguez2017b}. However, our approach differs in two
important ways---it operates at the quantum speed limit and assumes
controllability only for the system, not the bath.

Even when correlations are not created on purpose, they emerge inevitably when
components are coupled. This is ignored in theoretical proposals that assume
factorizing initial conditions.
Executing time-optimal qubit reset with and without artificially engineered
initial correlations would allow for an experimental comparison between
factorizing and non-factorizing initial conditions. This would be an important
step towards a better understanding of open quantum systems.

Enhancement of initial correlations by use of an ancialla or defect provides
a fresh perspective onto quantum reservoir engineering~\cite{PoyatosPRL96}. So
far, protocols for quantum reservoir engineering have targeted the creation of
non-trivial quantum states as steady state of some driven-dissipative dynamics,
see e.g.~\cite{PoyatosPRL96, PielawaPRL07, DiehlNatPhys08, KastoryanoPRL11},
assuming the evolution to be Markovian and the coupling to the environment to be
weak. While we have found non-Markovianity per se not to be relevant for the
success of qubit reset, we show that strong coupling to an engineered
environment allows for faster protocols and the emerging correlations to be
useful for a further speed up of the evolution. This suggests to explore quantum
reservoir engineering in scenarios beyond the weak coupling and Markov
approximations.

\begin{acknowledgments}
  We thank Jukka Pekola, Mikko M\"ott\"onen, Ronnie Kosloff, Pietro
  Liuzzo-Scorpo, Stefan Filipp, Felix Motzoi, and Kondra Tulja Varun for helpful
  discussions.
  Financial support from the Volks\-wagen\-stiftung, the Center of
  Quantum Engineering at Aalto University, the Academy of Finland (project no.
  287750), DAAD/Academy of Finland mobility grants, Agence nationale de la
  recherche (grant no. ANR-15-CE30-0023-01), and the PICS program of the CNRS is
  gratefully acknowledged. This work was done in part with the support of the
  Technische Universit\"at M\"unchen---Institute for Advanced Study, funded by
  the German Excellence Initiative and the European Union Seventh Framework
  Programme under grant agreement 291763.
\end{acknowledgments}

\begin{appendix}
\section{Influence of the Counter Rotating Terms} \label{app:RWA}
For obtaining analytical results, we need to employ the rotating wave
approximation (RWA). In the following, we therefore examine the influence of the
counter rotating terms in the interaction Hamiltonian~\eqref{eq:model:hamQT}. It
can be rewritten,
\begin{equation}
  \begin{aligned}
    \op{H}_{\text{int}}
    &=
    J \left(%
          \op{\sigma}^{+}_{\text{Q}} \otimes \op{\sigma}^{+}_{\text{TLS}}
        + \op{\sigma}^{+}_{\text{Q}} \otimes \op{\sigma}^{-}_{\text{TLS}}
        \right. \\ & \qquad \left.
        + \op{\sigma}^{-}_{\text{Q}} \otimes \op{\sigma}^{+}_{\text{TLS}}
        + \op{\sigma}^{-}_{\text{Q}} \otimes \op{\sigma}^{-}_{\text{TLS}}
      \right),
  \end{aligned}
\end{equation}
where $\op{\sigma}^{-}$ ($\op{\sigma}^{+}$) are the usual lowering (raising)
operators for two-level systems. The counter rotating terms are given by
$\op{\sigma}_{\text{Q}}^{+} \otimes \op{\sigma}_{\text{TLS}}^{+}$ and
$\op{\sigma}_{\text{Q}}^{-} \otimes \op{\sigma}_{\text{TLS}}^{-}$; they
are often neglected as part of a RWA\@. As we will show, these terms contribute
to the dynamics, i.e., the RWA is not a good approximation here. Nevertheless,
they have only a minor influence on the solution of the reset control problem.

\begin{figure*}[tb]
  \centering
  \includegraphics[width=0.99\linewidth]{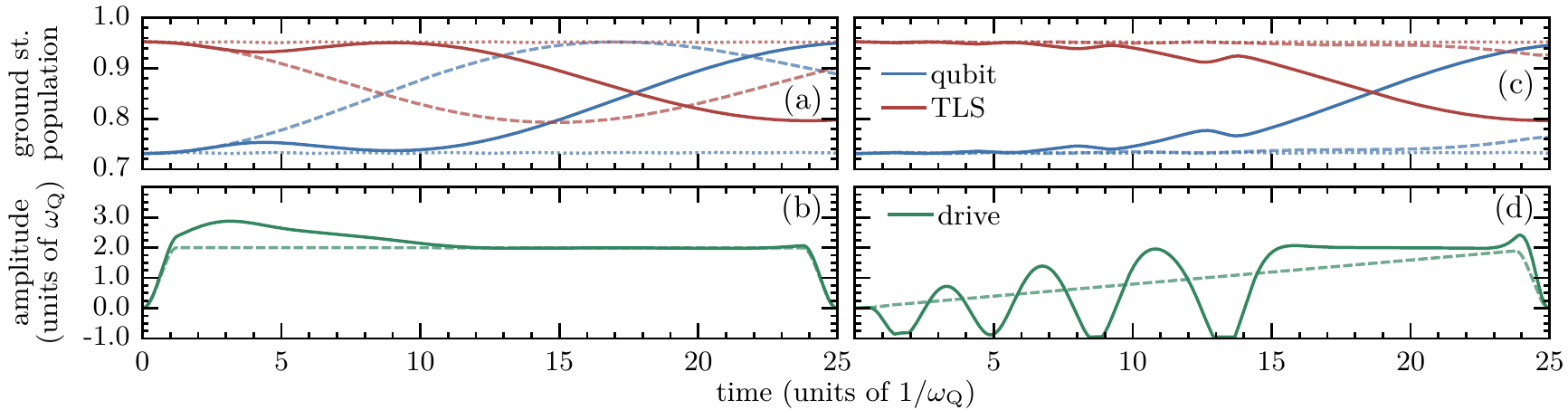}
  \caption{%
    Identical to Fig.~\ref{fig:dyn_fact} but employing the
    RWA~\eqref{eq:num:ham_int_RWA} for dynamics and optimizations. For the final
    errors, we find $\error_{T} = 5.01\%$ (a,b) and $\error_{T} = 5.41\%$ (c,d).
  }
  \label{fig:dyn_fact_RWA}
\end{figure*}
In the RWA, the interaction Hamiltonian becomes
\begin{align} \label{eq:num:ham_int_RWA}
  \op{H}_{\text{int}}^{\text{RWA}}
  &=
  J \left(%
        \op{\sigma}^{+}_{\text{Q}} \otimes \op{\sigma}^{-}_{\text{TLS}}
      + \op{\sigma}^{-}_{\text{Q}} \otimes \op{\sigma}^{+}_{\text{TLS}}
    \right).
\end{align}
Repeating the optimizations for the factorizing initial state
\eqref{eq:model:state_fact_init} under the RWA yields errors that are slightly
smaller ($\error_{T} = 5.01\%$ in Fig.~\ref{fig:dyn_fact_RWA}(a,b) and
$\error_{T} = 5.41\%$ in Fig.~\ref{fig:dyn_fact_RWA}(c,d)), compared to the case
when the counter-rotating terms are included ($\error_{T} = 5.04\%$ in
Fig.~\ref{fig:dyn_fact}(a,b) and $\error_{T} = 5.44\%$ in
Fig.~\ref{fig:dyn_fact}(c,d)). Employing the optimized fields from
Fig.~\ref{fig:dyn_fact_RWA} in the dynamics including the counter rotating terms
(without further optimization) results in only slightly increased final errors
$\error_{T} = 5.12\%$ (a,b) and $\error_{T} = 5.49\%$ (c,d). The errors are thus
affected only in the third digit, despite the dynamics and optimized fields in
Figs.~\ref{fig:dyn_fact} and~\ref{fig:dyn_fact_RWA} being visibly different.

In order to repeat this analysis for non-factorizing initial states, we have to
adjust the joint thermal state \eqref{eq:model:th_bipartite} of qubit and TLS\@,
\begin{align} \label{eq:num:th_bipartite_RWA}
  \op{\rho}^{\text{init}}_{2,\text{RWA}}
  =
  \op{\rho}^{\text{th}}_{\text{RWA}}
  =
  \frac{1}{Z}
  \begin{pmatrix}
    e^{\phi} & 0 & 0 & 0
    \\
    0 & \lambda + \frac{\delta}{\Omega} \mu & - \frac{2 J}{\Omega} \mu & 0
    \\
    0 & - \frac{2 J}{\Omega} \mu & \lambda - \frac{\delta}{\Omega} \mu & 0
    \\
    0 & 0 & 0 & e^{-\phi}
  \end{pmatrix},
\end{align}
where $\lambda = \cosh(x)$, $\mu = \sinh(x)$ and partition function $Z
= 2 \cosh\left(\phi\right) + 2 \cosh\left(x\right)$ with
\begin{equation}
  \begin{alignedat}{3}
    \delta
    &=
    \omega_{\text{Q}} - \omega_{\text{TLS}},
    \qquad
    &&\phi
    &&=
    \frac{\left(\omega_{\text{Q}} + \omega_{\text{TLS}}\right) \beta}{2},
    \\
    x
    &=
    \frac{\Omega \beta}{2},
    \qquad
    &&\Omega
    &&=
    \sqrt{\delta^{2} + 4 J^{2}}.
  \end{alignedat}
\end{equation}
The optimized final error in the RWA becomes $\error_{T} = 4.73\%$, compared to
$\error_{T} = 4.74\%$ in Fig.~\ref{fig:dyn_corr}(a,b,c). Using the RWA-optimized
field in the dynamics including the counter-rotating terms increases the final
error to only $\error_{T} = 4.78\%$. This is particularly remarkable, since not
only the interaction Hamiltonians differ, but also the initial states, cf.
Eqs.~\eqref{eq:model:th_bipartite} and \eqref{eq:num:th_bipartite_RWA}.
Similarly, for very strong initial correlations, we find $\error_{T} = 10.4\%$,
compared to $\error_{T} = 10.5\%$ in Fig.~\ref{fig:dyn_corr}(d,e,f); and use of
the RWA-optimized field in dynamics with the counter rotating terms increases
the error to only $\error_{T} = 10.6\%$. Similarly, we find our analysis of the
quantum speed limit and minimal achievable error in Fig.~\ref{fig:corr_var} to
be essentially independent of the RWA.

The small increase of the errors when using the RWA-optimized fields in dynamics
that include the counter-rotating terms is explained by larger final residual
correlations. However, the increase due to the counter-rotating terms is of the
order of $10^{-4}$, whereas all final errors quoted above correspond to residual
correlations of the order of $10^{-3}$. Overall, the increase is thus
negligible, and we conclude that the counter-rotating terms, while modifying the
dynamics, have no relevant influence on the achievable final error or, in other
words, the controllability of the problem. This has two important implications:
First, in order to identify control solutions for the reset problem, it is
sufficient to consider the interaction Hamiltonian in the
RWA~\eqref{eq:num:ham_int_RWA}. This will allow an analytical treatment, see
Sec.~\ref{sec:ana} below. Moreover, from an experimental perspective, a loss of
fidelity in the third digit is irrelevant and it might actually be advantageous
to use RWA-optimized fields, since these are generally much smoother, cf.
Fig.~\ref{fig:dyn_fact}(b,d) and Fig.~\ref{fig:dyn_fact_RWA}(b,d).

\section{Variable transformations} \label{app:derivation}
The RWA-Hamiltonian, neglecting counter rotating terms, reads
\begin{equation}
  \begin{aligned}
    \op{H}^{\text{RWA}}(t)
    &=
    \op{H}_{\text{Q}}(t) \otimes \uop_{\text{TLS}} + \uop_{\text{Q}}
    \otimes \op{H}_{\text{TLS}} + \op{H}^{\text{RWA}}_{\text{int}}
    \\
    &=
    \op{H}_{0}(t) + \op{H}^{\text{RWA}}_{\text{int}},
  \end{aligned}
\end{equation}
with $\op{H}^{\text{RWA}}_{\text{int}}$ defined as in
Eq.~\eqref{eq:num:ham_int_RWA}. Performing a unitary transformation with
transformation operator
\begin{align}
  \op{O}(t)
  =
  \text{exp}\left\{- \im \op{H}_{0}^{\text{RWA}}(t) t\right\}
\end{align}
yields a transformed state $\op{\rho}'(t)$ and Hamiltonian $\op{H}'(t)$,
\begin{equation}
  \begin{aligned}
    \op{\rho}'(t)
    &=
    \op{O}^{\dagger}(t) \op{\rho}(t) \op{O}(t),
    \\
    \op{H}'(t)
    &=
    \op{O}^{\dagger}(t) \op{H}(t) \op{O}(t) - \im \op{O}^{\dagger}(t)
    \frac{\text{d} \op{O}(t)}{\text{d}t}.
  \end{aligned}
\end{equation}
This yields the Liouville-von Neumann equation~\eqref{eq:ana:LvN_rotating}.
Starting from there, we summarize in the following the variable transformations
required to derive Eqs.~\eqref{eq:ana:veceqs_S1} and~\eqref{eq:ana:veceqs_S2} in
Section~\ref{subsec:ana:geo_eq}. First, we represent the density matrix in the
rotating frame, $\op{\rho}'(t)$, in terms of $16$ real variables, $x_i(t)\in
\mathbb{R}$, dropping the explicit time-dependence for all quantities in the
following,
\begin{align} \label{eq:ana:rho}
  \op{\rho}'
  =
  \begin{pmatrix}
    x_{1} & x_{5} + \im x_{6} & x_{7} + \im x_{8} & x_{9} + \im x_{10}
    \\
    x_{5} - \im x_{6} & x_{2} & x_{11} + \im x_{12} & x_{13} + \im x_{14}
    \\
    x_{7} - \im x_{8} & x_{11} - \im x_{12} & x_{3} & x_{15} + \im x_{16}
    \\
    x_{9} - \im x_{10} & x_{13} - \im x_{14} & x_{15} - \im x_{16} & x_{4}
  \end{pmatrix}\,.
\end{align}
The set $\left\{x_{1}, \dots, x_{16}\right\}$ spans the entire state space, and
the equation of motion~\eqref{eq:ana:LvN_rotating} becomes
\begin{subequations}\label{eq:ana:control_system_x}
\begin{align}
  \dot{\vec x}
  =
  J_{1} \vec f_{1}(\vec x) + J_{2} \vec f_{2}(\vec x)
  + \alpha \vec f_{3}(\vec x),
\end{align}
with $\vec x = \left(x_{1},\dots,x_{16}\right)^{\top}$,
\begin{align}
  \vec f_{1}
  =
  \begin{pmatrix}
    0 \\
    - 2 x_{12} \\
    2 x_{12} \\
    0 \\
    - x_{8} \\
    x_{7} \\
    - x_{6} \\
    x_{5} \\
    0 \\
    0 \\
    0 \\
    x_{2} - x_{3} \\
    x_{16} \\
    - x_{15} \\
    x_{14} \\
    - x_{13}
  \end{pmatrix},
  \quad
  \vec f_{2}
  =
  \begin{pmatrix}
    0 \\
    - 2 x_{11} \\
    2 x_{11} \\
    0 \\
    - x_{7} \\
    - x_{8} \\
    x_{5} \\
    x_{6} \\
    0 \\
    0 \\
    x_{2} - x_{3} \\
    0 \\
    - x_{15} \\
    - x_{16} \\
    x_{13} \\
    x_{14}
  \end{pmatrix},
  \quad
  \vec f_{3}
  =
  \begin{pmatrix}
    0 \\
    0 \\
    0 \\
    0 \\
    0 \\
    0 \\
    2 x_{8} \\
    - 2 x_{7} \\
    2 x_{10} \\
    - 2 x_{9} \\
    2 x_{12} \\
    - 2 x_{11} \\
    2 x_{14} \\
    - 2 x_{13} \\
    0 \\
    0
  \end{pmatrix}.
\end{align}
\end{subequations}
and $J_{1}$, $J_{2}$ and $\alpha$ given in Eq.~\eqref{eq:alpha}.
The vector fields $\vec f_{1}(\vec x)$
$\vec f_{2}(\vec x)$ and $\vec f_{3}(\vec x)$ govern the admissible directions
for the evolution of the state $\vec x$, whereas $J_{1}$, $J_{2}$ and $\alpha$
determine their relative magnitude for each direction. With the
representation~\eqref{eq:ana:rho}, the purity of the qubit becomes
\begin{equation}
  \begin{aligned}
    \mathcal{P}_{\text{Q}}
    &=
    \left(x_{1} + x_{2}\right)^{2}
    +
    \left(x_{3} + x_{4}\right)^{2}
    \\
    &\quad +
    2 \left(x_{7} + x_{13}\right)^{2}
    +
    2 \left(x_{8} + x_{14}\right)^{2}.
  \end{aligned}
\end{equation}

The set of the coupled equations~\eqref{eq:ana:control_system_x} is separated in
two disjunct sets by introducing new variables, $z_i\in \mathbb{R}$. The
relevant ones are given by
\begin{equation} \label{eq:app:z_transform}
  \begin{alignedat}{2}
    z_{1} &= x_{1} + x_{2} - 1/2,
    \qquad
    &&z_{5} = x_{7} + x_{13},
    \\
    z_{2} &= x_{12},
    \qquad
    &&z_{6} = x_{6} - x_{16},
    \\
    z_{3} &= x_{11},
    \qquad
    &&z_{7} = x_{8} + x_{14},
    \\
    z_{4} &= - 2 x_{1} - x_{2} - x_{3},
    \qquad
    &&z_{8} = x_{5} - x_{15}.
  \end{alignedat}
\end{equation}
There are eight further variables, $z_{9}, \dots, z_{16}$, that are required to
span the entire state space. However, these variables are not coupled to $z_{1},
\dots, z_{8}$, so they can be ignored for the maximization of the purity.

Using the new variables and exploiting that
$\tr{}{}{\op{\rho}'}=x_{1}+x_{2}+x_{3}+x_{4}=1$, the qubit purity simplifies to
Eq.~\eqref{eq:ana:qubit_purity_z}. Moreover, the equations of motion for $z_{1},
\dots, z_{8}$ decouple into two independent subspaces. One subspace is $S_{1}
= \left\{z_{1}, z_{2}, z_{3}\right\}$ with the equations of motion given in
Eq.~\eqref{eq:ana:veceqs_S1}, where $z_{1}^{\text{c}} = -\left(z_{4}
+ 1\right)/2$ is a constant since $\dot{z}_{4} = 0$. The other subspace is
$S_{2} = \left\{z_{5}, z_{6}, z_{7}, z_{8}\right\}$ with the equations of motion
given by Eq.~\eqref{eq:ana:veceqs_S2}.

\end{appendix}


%

\end{document}